# Small-Angle Scattering Studies of Intrinsically Disordered Proteins and their Complexes

Tiago N. Cordeiro[a], Fátima Herranz-Trillo[a,c], Annika Urbanek[a], Alejandro Estaña[a,b], Juan Cortés[b], Nathalie Sibille[a], Pau Bernadó[a,*] a-Centre de Biochimie Structurale. INSERM U1054, CNRS UMR 5048, Université de Montpellier. 29, rue de Navacelles, 34090-Montpellier, France.
b- LAAS-CNRS, Université de Toulouse, CNRS, Toulouse, France.
c- Department of Pharmacy and Department of Drug Design and Pharmacology, University of Copenhagen, Universitetsparken 2, 2100 Copenhagen, Denmark.

**Contact Author:** Pau Bernadó (pau.bernado@cbs.cnrs.fr). Tel. +33 467417705

**Abstract:**

Intrinsically Disordered Proteins (IDPs) perform a broad range of biological functions. Their relevance has motivated intense research activity seeking to characterize their sequence/structure/function relationships. However, the conformational plasticity of these molecules hampers the application of traditional structural approaches, and new tools and concepts are being developed to address the challenges they pose. Small-Angle Scattering (SAS) is a structural biology technique that probes the size and shape of disordered proteins and their complexes with other biomolecules. The low-resolution nature of SAS can be compensated with specially designed computational tools and its combined interpretation with complementary structural information. In this review, we describe recent advances in the application of SAS to disordered proteins and highly flexible complexes and discuss current challenges.

**Introduction**

In the last two decades, Intrinsically Disordered Proteins or Regions (IDPs/IDRs) have emerged as fundamental molecules in a broad range of crucial biological functions such as cell signaling, regulation, and homeostasis [1,2,3**]. Due to their lack of a permanent secondary and tertiary structure, IDPs and IDRs are highly plastic and have the capacity to perform specialized functions that complement those of their globular (folded) counterparts [4]. Disordered regions, which can finely adapt to the structural and chemical features of their partners, are very well suited for protein-protein interactions and are thus abundant in hub positions of interactomes [5,6,7].

The importance of disordered proteins in a multitude of biological processes has fostered intense research efforts that seek to unravel the structural bases of their function. Nuclear Magnetic Resonance (NMR) has been the main structural biology technique used to characterize the conformational preferences at residue level, and, therefore, to localize partially structured elements [8,9]. However, a number of structural features related to the overall size and shape of IDPs or their complexes remain elusive to NMR. To study these properties, thereby complementing NMR residue-specific information, Small-Angle Scattering (SAS) of X-rays (SAXS) or Neutrons (SANS) is the most appropriate technique [10,11,12]. Although SAS is a low-resolution technique, the data obtained is sensitive to large-scale protein fluctuations and the presence of multiple species and/or conformations in solution [13,14,15]. However, the conversion of SAS properties into structural restraints is challenging due to the enormous conformational variability of IDPs and the ensemble-averaged nature of the experimental data [16]. The quantitative analysis of these data in terms of structure has prompted the development of computational approaches to both model disordered proteins and to use ensembles of conformations to describe the experimental data. Here we highlight the most relevant developments and applications of SAS to IDPs and IDRs, with a special emphasis on the computational strategies required to fully exploit the data in order to achieve biologically insightful information.

**Structural models of IDPs and their experimental validation**

For disordered proteins, the structural insights gained from overall SAS parameters, such as the radius of gyration, $R_g$, the pairwise intramolecular distance distribution, $p(r)$, and the maximum intramolecular distance, $D_{max}$, are limited. Neither these parameters nor the traditional Kratky representation, $I(s)s^2$ *vs s*, which qualitatively report on the compactness of biomolecules in solution, directly account for the ensemble nature of disordered proteins. In order to fully exploit the structural and dynamic information encoded in SAS data, it is necessary to use realistic three-dimensional (3D) models. However, the generation of conformational ensembles of disordered proteins is extremely challenging, mainly because of the flat energy landscape and the large number of local minima separated by low-energy barriers [17]. The most popular methods to generate 3D models of IDPs are based on residue-specific conformational landscapes derived from large databases of crystallographic structures [18,19,20*]. However, the main limitation of these approaches is the absence of sequence context information, thereby precluding the prediction of transiently formed secondary structure elements or the presence of long-range interactions between distant regions of the protein. Accurate energy models (force-fields) accounting for the interactions within the chain and with the solvent are required to describe these features. The development of specific force-fields to study conformational fluctuations in disordered proteins is a very active field of research [21,22,23,24]. Molecular Dynamics (MD) or Monte-Carlo (MC) simulations, when an appropriate energy description is provided, are suitable methods to correctly sample the conformational space of IDPs. However, the high-dimensionality and the breadth of the energy landscape hamper exhaustive exploration of this space. Replica Exchange MD (REMD) [25,26], which exchanges conformations between parallel simulations running at multiple temperatures, or Multiscale Enhanced Sampling (MSES) [27], which couples temperature and Hamiltonian replica exchange, have been proposed to enhance the conformational exploration of MD methods. The performance of MD-based methods can also be improved by the inclusion of experimental data to delimit the exploration to the most relevant regions of the conformational space [28,29,30].

The quality of computational models of disordered proteins is normally validated using experimental data. The $R_g$ derived from the low-angle region of SAXS curves or from the $p(r)$ function is an excellent probe of the overall size of a particle in solution. $R_g$ compilations have been extensively used to validate models of denatured and natively disordered proteins through Flory's relationship, which correlates the $R_g$ observed with the number residues of the chain [31,14]. The compilation of the $R_g$s from 76 IDPs (Figure 1) reveals that these proteins are more compact than chemically denatured ones. It has been shown that denatured proteins present an enhanced sampling of extended conformations, probably due to the interaction of the protein with chemical agents [32]. Importantly, deviations from the expected $R_g$ values for canonical random-coil behavior, which is represented by the green line in figure 1, indicate the presence of structural features that modify the overall size of the particle in solution towards more extended or more compact (Figure 1). The extendedness detected using this analysis for several Tau protein constructs has been linked to the presence of secondary structural elements probed by NMR [33]. These structural properties can be more thoroughly examined when the complete SAXS curve is used to validate the ensemble models of peptides [34] or proteins [19,35,36].

**Ensemble approaches**

In the last decade, ensemble methods have become highly popular to structurally characterize disordered proteins. Guided by experimental data, these methods aim to derive accurate ensemble models of flexible proteins. Several strategies that apply these methods to SAS data have been reported: Ensemble Optimization Method (EOM) [37,38]; Minimal Ensemble Search (MES) [39]; Basis-Set Supported SAXS (BSS-SAXS) [40]; Maximum Occurrence (MAX-Occ) [41]; Ensemble Refinement of SAXS (EROS) [42]; Broad Ensemble Generator with Re-weighting (BEGR) [43]; and Bayesian Ensemble SAXS (BE-SAXS) [44]. These methods share a common strategy that consists of the following three consecutive steps: (i) computational generation of a large ensemble that describes the conformational landscape of the protein; (ii) calculation of the theoretical SAXS curves from the individual conformations; and (iii) use of a multiparametric optimization method to select a sub-ensemble of conformations that collectively describe the experimental profile. Despite the common strategy, these approaches present distinct features in the three steps. Readers are referred to the original articles for detailed descriptions. The availability of ensemble methods has transformed the study of flexible proteins by SAS. Ensemble methods provide a description in terms of the statistical distributions of structural parameters or conformations that is revolutionary with respect to traditional analyses based on averaged parameters extracted from raw data. Using this power, structural perturbations exerted by temperature [45*,46], buffer composition [47], or mutations [48] have been monitored in terms of ensembles of conformations.

Despite the popularity of ensemble methods, several aspects are still under debate. The most relevant ones are the use of discrete descriptions for entities that probe an astronomical number of conformations, and the statistical significance of ensembles derived from data containing a very limited amount of information. The strategies described use distinct philosophies to address these issues, including the search for the minimum number of conformations to describe the data [38,39], the representation of the optimal solution as a distribution of low-resolution structural parameters such as $R_g$ or $D_{max}$ [37], and the application of Bayesian statistics [40,44] or maximum entropy approaches [42]. Regardless of the strategy used to derive an ensemble of conformations compatible with the experimental data, one must be careful on the structural interpretation of the final solution. The optimized ensemble is a representation of the behavior of the protein in solution and not the exact enumeration of the conformations adopted by the protein. Consequently, the final ensemble can only be used to derive structural features that describe the protein. Importantly, the nature of these features depends on the experimental data used to derive the model. If only SAS data have been used, then an assessment of the degree of flexibility, and the size and shape distributions sampled by the protein can be obtained from the ensemble. Conversely, conformational preferences at residue level can be extracted if NMR information probing structure in a residue-specific manner is used along the refinement.

**Enriching the definition of conformational ensembles of IDPs with complementary information**

The definition of protein ensembles derived from SAS data using ensemble methods is limited to the overall structure and the space sampled by the protein in solution. Although this is an important improvement with respect to classical approaches, several crucial features, such as the localization of secondary structural elements or compact regions, remain elusive using this approach. Considerable research efforts have been channeled into enriching the resolution of the resulting ensemble with complementary information.

NMR is the only technique that can provide atomic-resolution information on IDPs and, consequently, it is the most common method applied in combination with SAS [49]. NMR is highly versatile and can measure multiple observables reporting on protein structure and dynamics [50]. Concretely, information reporting on the backbone conformational preferences at residue level can be probed by means of time- and ensemble-averaged chemical-shifts (CSs), J-couplings and Residual-Dipolar Couplings (RDCs). NMR can also probe long-range interactions within a protein chain or in protein complexes through Paramagnetic Relaxation Enhancement (PRE) experiments. In these experiments, a stable radical or a paramagnetic metal is introduced in a specific position of the chain, and the spatially close atoms can be identified by a decrease in their signal intensity that is proportional to the distance.

The best manner to exploit the complementarity between NMR and SAS is to integrate the experimental data into the same refinement protocol. The programs ENSEMBLE [51,52] and ASTEROIDS [53] derive ensembles of disordered proteins by collectively describing SAXS curves, in addition to several NMR observables. These powerful approaches seek to find the appropriate way to combine data with very different information content while avoiding overfitting. In a pioneering study, ensembles of Tau and α-synuclein were determined by combining SAXS with multiple backbone CS, RDC, and PRE datasets [54**]. Those authors addressed the optimal combination of experimental data and the overfitting problem with extensive cross-validation tests that substantiated conformational bias in the aggregation-nucleation regions for both proteins.

Other structural techniques such as single molecule Fluorescence Resonance Energy Transfer (smFRET) [55] and Electron Paramagnetic Resonance (EPR) [56,57**] have been combined with SAXS to study large and flexible complexes. Recent developments in Mass Spectrometry (MS) offer novel sources of structural information [58]. Ion Mobility Spectrometry (IMS) can capture, in a similar way to SAS, the overall properties of conformational ensembles of disordered proteins. However, a recent study comparing IMS and SAXS data for some IDPs suggests that the conformations sampled in solution and in gas-phase are not equivalent [59]. Hydrogen/Deuterium Exchange MS (HDX/MS) probes structural elements in proteins by identifying regions that are protected from the exchange with solvent protons [58]. The availability of fast HDX/MS methods enables the exploration of secondary structural elements in IDPs and localizing their interaction sites with globular partners [60]. In a recent study HDX/MS information was combined with SAXS to study the calcium-induced structure formation in RD, a protein hosting repeated regions able to bind this cation [61].

The structural definition of a SAXS derived ensemble model can also be enriched by the simultaneous analysis of curves measured for multiple deletion mutants of the same IDP [37]. When applied to two different isoforms of Tau protein, this approach identified the repeat region of the protein as the origin of distinct global rearrangements of its flanking regions [62].

The large toolbox of structural techniques that can probe distinct structural features of IDPs will result in a better understanding on their structure-function relationship. In this regard, the future development of robust and reliable ways to integrate biophysical measurements in ensemble approaches is imperative when addressing complex biomolecular entities such as IDPs and their complexes.

**Disordered proteins in complexes**

The biological function of many IDPs is manifested when they recognize their biological folded partners [5]. This recognition frequently involves linear motifs of the disordered chain, which, upon binding, adopt relatively fixed conformations while the rest of the IDP remains flexible [63].

The relevance of protein-protein complexes involving disordered partners has promoted growing interest in unraveling their structural characterization, with the aim to understand the bases of their biological activity. This structural characterization is complex and poses multiple challenges to traditional structural biology methods. SAXS has emerged as a valuable alternative. However, overall structural parameters or *ab initio* reconstructions derived from SAXS curves cannot capture the inherent plasticity of these complexes [64,65*,66]. Hybrid (or integrative) methods that combine information from multiple techniques, thus exploiting their individual strengths, are the most appropriate approaches to study highly flexible complexes [67]. In this context, it is important to describe how different structural biology techniques probe complexes involving IDPs (Figure 2). Due to the dynamic nature of the interaction and the distinct hydrodynamic properties of the globular and disordered parts of the complex, NMR generally detects only those regions that remain flexible upon binding. Although not general, it is sometimes possible to crystallize the globular partner in the presence of a small peptide corresponding to the interacting region of the IDP. Therefore, X-ray crystallography provides an atomic resolution picture of the interacting regions that is complementary to NMR since the two techniques probe non-overlapping parts of the same entity [68]. Conversely, SAXS probes the complete assembly and can be used to integrate the information from both NMR and X-ray crystallography. If one of the partners is deuterated, contrast variation SANS experiments can be performed and the individual components of the assembly can be alternatively highlighted depending on the $D_2O/H_2O$ ratio of the buffer. The power of combining multiple techniques is exemplified in the study of the interaction of the Vesicular Stomatitis Virus (VSV) nucleoprotein ($N^0$) and the dimeric phosphoprotein (P), a high-affinity complex that precludes the oligomerization of $N^0$ *in vivo* [69**]. Using EOM, the authors simultaneously fitted one SAXS curve and four SANS curves measured at different contrast levels for the complex of $N^0$ with deuterated P protein. The additional information provided by the distinct contribution of the two proteins in the SANS experiments notably improved the description of the conformational properties of the complex.

In many cases, the conformational mobility of the interacting region of the IDP is reduced (or frozen) upon binding to the biological partner. There is an entropic cost associated with this rigidification that often leads to low- to moderate-affinity complexes ($K_d$ > 1 μM) [63]. The structural modulation of the affinity is key to achieving tunable responses to external signals, thereby explaining the prevalent role of disordered proteins in signaling processes [2,3**]. In the concentration range normally used in SAXS experiments, the complex is in equilibrium with the free forms of the two partners, thereby giving rise to population-weighted averaged SAXS curves (Figure 3A). This scenario can be even more complex if one or both of the partners have multiple equivalent or similar binding sites (Figure 3B,C). In this case, the polydispersity of the mixture increases as a result of the presence of several complexes with distinct stoichiometries.

The interpretation of SAS data from polydisperse samples is challenging [70]. Although the coupling of SAXS to Size-Exclusion Chromatography (SEC-SAXS) can, in some instances, separate the components of the mixture, there are multiple examples where the coexistence of multiple species is unavoidable. In these circumstances and with the aim to isolate the contribution of the individual species within complex mixtures, analytical approaches have been developed to decompose large SAXS titration datasets [71,72]. This decomposition is easier when prior structural knowledge of the species is used for the analysis [70]. However, to apply this strategy to low-affinity flexible complexes, accurate conformational descriptions of all species in the free and bound forms are mandatory. The analysis of SAS data measured in samples with different relative concentrations of both partners seems the most appropriate strategy to enrich the information content in order to structurally characterize these extremely challenging scenarios (Figure 3).

**Conclusions and outlook**

During the last decade, SAS has been added to the toolbox of techniques used to study conformational fluctuations in proteins. This dynamic revolution of SAS is linked to the development of computational tools able to describe the conformational landscape of biomolecules and ensemble approaches with the capacity to interpret SAS data in terms of structural variability. These computational tools, which use chemical and structural knowledge of biomolecules, partially compensate for the limited amount of information coded in a SAS curve. Therefore, the capacity to fully exploit the structural information held in SAS data will necessarily be linked to the development of more advanced and precise computational approaches with specially developed force-fields. This notion is especially applicable to IDPs and IDRs, which populate a huge number of conformational states. For these proteins, SAS can be enriched with complementary information obtained by NMR, smFRET, EPR, or MS, and integrated into a common ensemble model embedding structure and dynamics. A particularly challenging subclass of IDPs is that containing Low-Complexity Regions (LCRs), which are involved in multitude of biological processes and are related to severe pathologies. LCRs are unusually simple protein sequences with a strong amino acid composition bias. The resulting similarity of chemical environments within their sequence hampers their structural characterization by NMR. SAS can be a valuable alternative through which to study this important but structurally neglected family of proteins [73,74,75].

The function of multitude of IDPs is determined by their interaction with biomolecular partners to form assemblies, which, in many cases, are of low to moderate affinity. The capacity of SAS to probe the size and shape of particles in solution places this technique in a unique position to address these polydisperse scenarios. A case in point is the fibrillation process that several IDPs undergo to form amyloids, which are linked to severe diseases. The decomposition of time-dependent SAXS datasets has been successfully used to characterize intermediate oligomeric forms [76*,77], thereby validating SAXS as a practical tool for this purpose

The need to understand the mechanisms underlying complex cellular processes and recent technical and conceptual advances in structural biology techniques across the board have prompted researchers to tackle challenging systems that were inaccessible some years ago. Many of these systems are inherently dynamic and/or polydisperse and can be exquisitely probed by SAS. As a consequence, we anticipate that SAS will take on greater relevance in hybrid approaches where its unique information will be synergistically integrated with data from multiple sources to deliver accurate structural and dynamic models of disordered proteins and their complexes.

**Conflict of interest statement**
Authors declare no conflict of interest

**Acknowledgements**
This work was supported by the ERC-CoG chemREPEAT, SPIN-HD – Chaires d'Excellence 2011 from the *Agence National de Recherche* (ANR), ATIP-Avenir, and the French Infrastructure for Integrated Structural Biology (FRISBI – ANR-10-INSB-05-01) to PB. FHT is supported by INSERM and the Sapere Aude Programme SAFIR of the University of Copenhagen. AU is supported by a grant from the *Fondation pour la Recherche Médicale*.

# References and recommended reading

**FIGURES:**

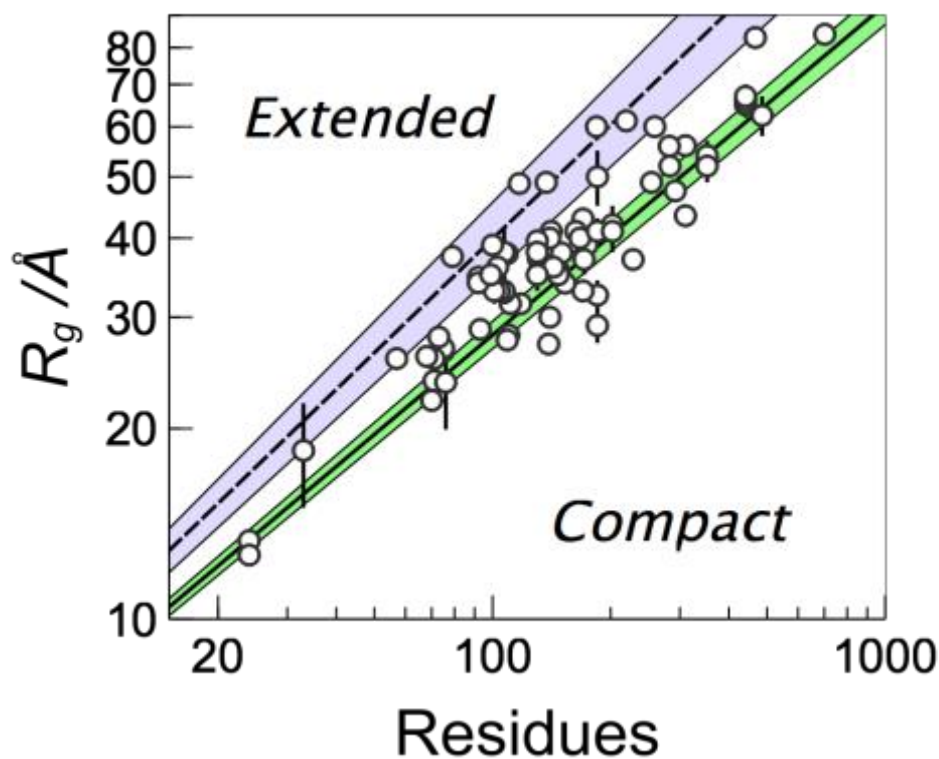

***Figure 1****. $R_g$ values from 76 IDPs as a function of the number of residues of the protein are plotted in Log-Log scale. Only proteins lacking a permanent secondary or tertiary structure were considered for the compilation. Proteins with ordered domains, molten globules, or denatured proteins were not considered. Straight lines correspond to Flory's relationships parametrized for denatured proteins using experimental data (purple-dashed) [31] and IDPs using computational ensembles calculated with Flexible-Meccano (green-solid) [32]. Colored bands correspond to uncertainty of the parametrization for both models. Some IDPs contain local structural features and consequently they are globally more extended or more compact than expected for a random coil. These structural features, even if transient, can be manifested in the experimental $R_g$.*

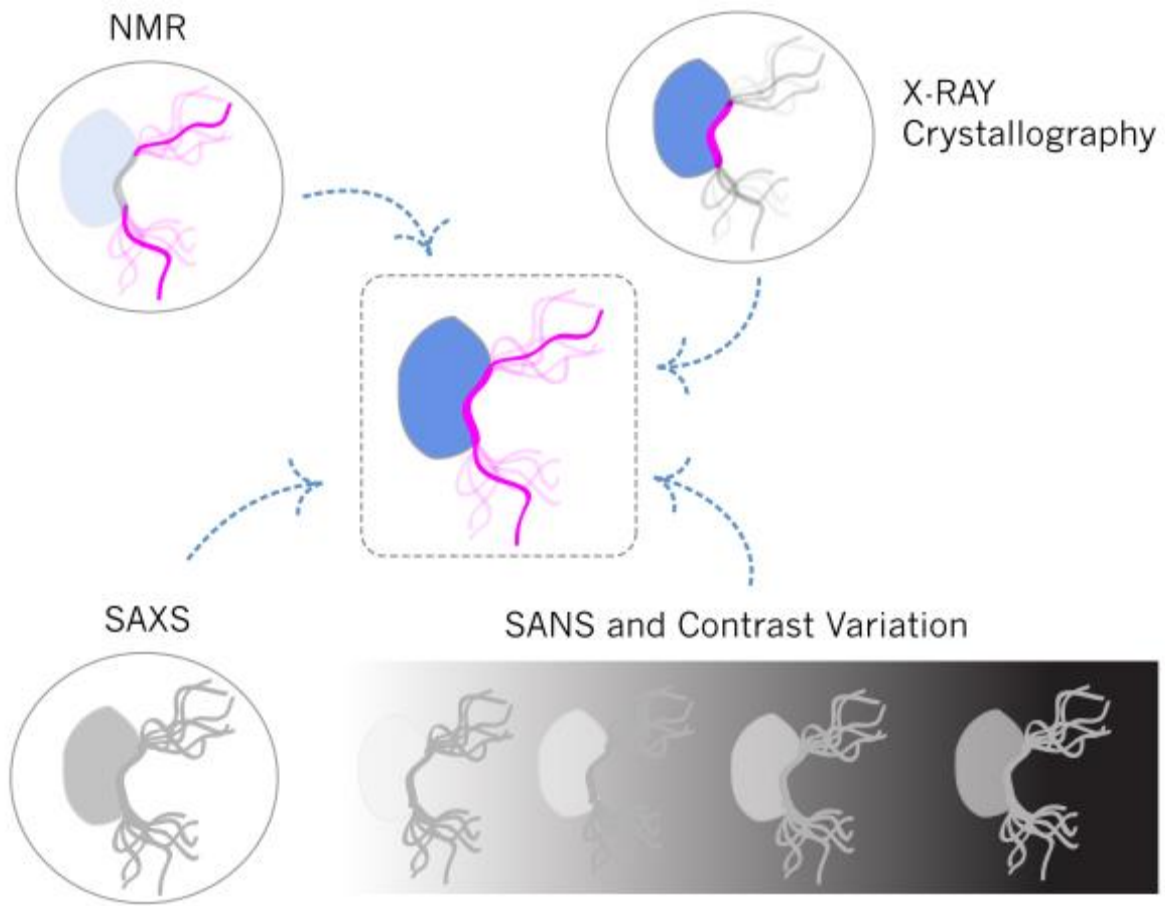

*Figure 2. Cartoons representing the structural sensitivity of NMR, X-ray crystallography, and SAS for a complex involving a disordered protein (central cartoon). NMR normally probes the flexible regions of these complexes while the globular partner and the interacting region remain invisible. Crystallography provides detailed information of the interacting region of the complex but not for the flexible parts. SAXS probes the complete ensemble, although the details cannot be assessed due to its inherent low-resolution. SANS, through contrast variation experiments, can probe independently both partners in the context of the complex depending on the deuteration level of the partners and the $D_2O/H_2O$ of the buffer. SAS is an ideal tool to integrate NMR and crystallographic information to build complete structural and dynamic models of disordered biomolecular complexes.*

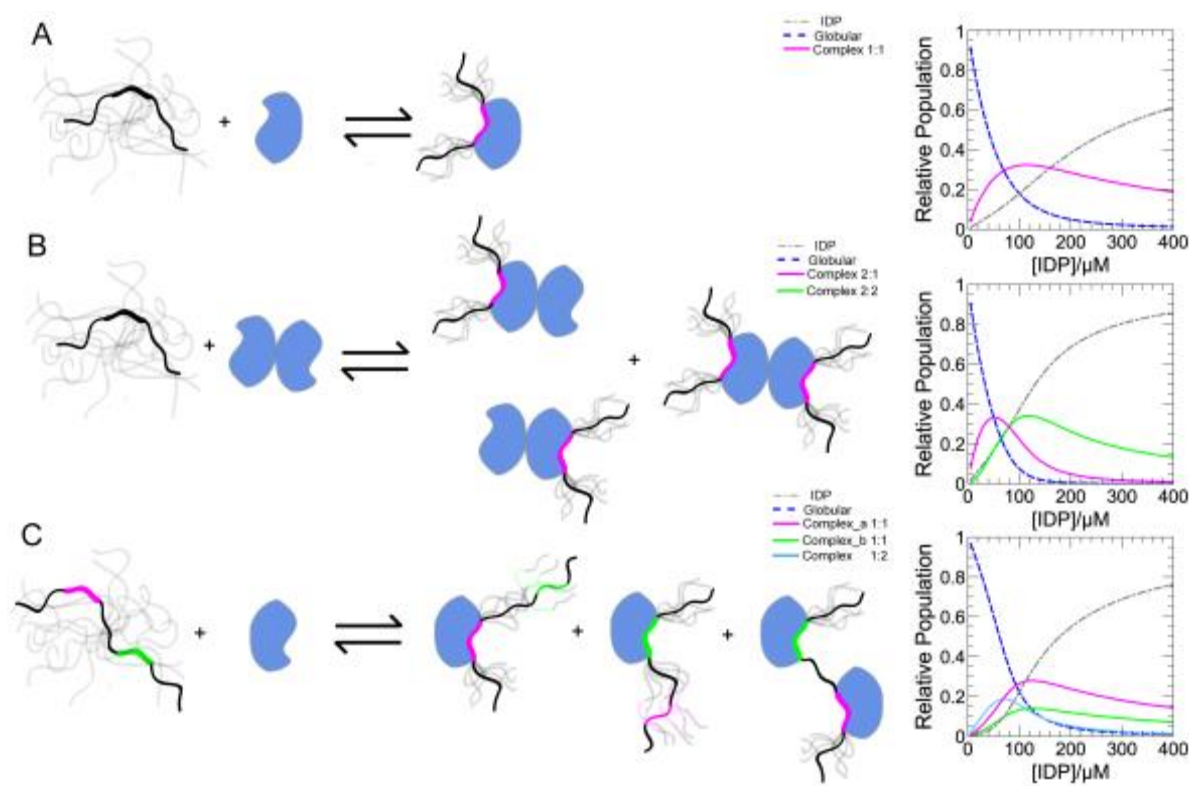

*Figure 3. Examples of polydisperse scenarios that can occur in low-affinity complexes involving an IDP and a globular partner. (A) Both proteins have a single binding site. The complex is in equilibrium with the free forms of both proteins. (B) The globular partner is a dimer and has two identical binding sites. The free forms are in equilibrium with three possible complexes recognizing one or two binding sites of the globular partner. Due to the symmetry of the dimer, the two singly bound complexes are however indistinguishable by SAS. (C) The IDP presents two similar binding sites (pink and green). The free forms are in equilibrium with two 1:1 complexes using a distinct IDP interacting site to bind the globular partner, and a complex where the IDP simultaneously interacts with two globular partners. On the right part of the figure, three panels are displayed representing the molar fraction of each species along a simulated titration experiment for each scenario. These populations were computed assuming a fixed concentration of the globular partner, [globular] = 100 µM, and increasing concentrations of IDP, [IDP], from 1 µM to 400 µM. A common dissociation constant $K_d$ = 20 µM was used for scenarios A and B, in panel C the two IDP binding sites, pink and green, display a $K_d$ = 20 µM and 40 µM, respectively. These panels exemplify the inherent polydispersity moderate affinity complexes, and how multiple titration experiments will probe differently the species present and their relative populations.*